\documentclass[pra,aps,10pt,superscriptaddress,twocolumn,floatfix]{revtex4-1}  
\usepackage{graphicx,color}
\usepackage{amsmath,amssymb,bm}
\usepackage{braket}		

\DeclareMathOperator{\Tr}{Tr}

\usepackage[plainpages=false,pdfpagelabels,colorlinks=true,linkcolor=red,urlcolor=blue,citecolor=blue,pdftitle={Titl}e,pdfauthor={},pdfdisplaydoctitle=true,pdfduplex=DuplexFlipLongEdge]{hyperref}

\definecolor{darkred}{rgb}{0.90,0.2,0.2}
\definecolor{darkgreen}{rgb}{0,0.60,.2}
\definecolor{darkblue}{rgb}{0.1,0.3,1}
\definecolor{grey}{cmyk}{0,0,0,0.25}
\definecolor{orange}{cmyk}{0,0.6,0.8,0}

\begin{document}
\title{Emergent eigenstate solution for generalized thermalization}

\author{Yicheng Zhang}
\affiliation{Department of Physics, The Pennsylvania State University, University Park, Pennsylvania 16802, USA}
\author{Lev Vidmar}
\affiliation{Department of Theoretical Physics, J. Stefan Institute, SI-1000 Ljubljana, Slovenia}
\affiliation{Department of Physics, Faculty of Mathematics and Physics, University of Ljubljana, SI-1000 Ljubljana, Slovenia}
\author{Marcos Rigol}
\affiliation{Department of Physics, The Pennsylvania State University, University Park, Pennsylvania 16802, USA}

\begin{abstract}
Generalized thermalization is a process that occurs in integrable systems in which unitary dynamics, e.g., following a quantum quench, results in states in which observables after equilibration are described by generalized Gibbs ensembles (GGEs). Here we discuss an emergent eigenstate construction that allows one to built emergent local Hamiltonians of which one eigenstate captures the entire generalized thermalization process following a global quantum quench. Specifically, we study the emergent eigenstate that describes the quantum dynamics of hard-core bosons in one dimension (1D) for which the initial state is a density wave and they evolve under a homogeneous Hamiltonian.
\end{abstract}
\maketitle

Much progresses has been made in the field of nonequilibrium quantum dynamics in the past two decades~\cite{polkovnikov_sengupta_review_11, eisert_friesdorf_review_15, dalessio_kafri_16, mori_ikeda_18, deutsch_18}. One of the focuses in this field has been understanding how to describe isolated quantum systems after equilibration. In addition to being of fundamental interest, this is of relevance to experiments with ultracold quantum gases~\cite{kinoshita_wenger_06, gring_kuhnert_12, trotzky_chen_12, meinert13, langen_erne_15, clos_porras_16, kaufman_tai_16, Neill2016, tang_kao_18, malvania2020}. Thanks to both experimental and theoretical studies, we now know that after equilibration observables in generic (nonintegrable) quantum systems can be described using traditional ensembles of statistical mechanics while in integrable systems they can be described using generalized Gibbs ensembles~\cite{rigol_dunjko_07, cazalilla_2006, calabrese_essler_2011, ilievski15, piroli_vernier_16, ilievski_quinn_17, piroli_vernier_17, pozsgay_vernier_17}. This is theoretically understood in the context of eigenstate thermalization for generic systems~\cite{deutsch_91, srednicki_94, rigol_dunjko_08, dalessio_kafri_16} and of generalized eigenstate thermalization for integrable systems~\cite{cassidy_clark_11, caux_essler_13, vidmar16}. The actual dynamics of observables between their initial values and the equilibrated ones is in general nonuniversal and needs to be studied in a case by case basis. 

A different focus in the field of nonequilibrium quantum dynamics has been the realization of exotic states of matter, specially those that may not be accessible in equilibrium. This is a topic on which periodic driving has been the main focus of attention~\cite{bukov_dalessio_review_15}, e.g., to realize nontrivial topological states~\cite{Oka2009, Kitagawa2010, Lindner2011, Rechtsman2013, Jotzu2014, dalessio_rigol_15}. A special class of quantum quenches (geometric quenches) has also been used to created exotic states in strongly interacting one-dimensional bosonic systems. For example, to produce dynamical quasicondensation at finite momentum~\cite{rigol04, vidmar15} as well as expanding bosonic gases with fermionic momentum distributions~\cite{rigol05, minguzzi05, wilson_20}. Surprisingly, the latter far-from-equilibrium states exhibit power-law correlations and low entanglement typical of gapless ground states in one dimension. Those states were recently shown to be eigenstates (and specifically ground states) of emergent local Hamiltonians~\cite{vidmar_iyer_17, vidmar_xu_17}.

In this work we tackle a question that bridges the two focuses mentioned above, namely, is it possible for an eigenstate of an emergent local Hamiltonian to describe the entire path to (generalized) thermalization after a global quantum quench? In contrast to the states generated by geometric quenches in Refs.~\cite{rigol04, vidmar15, rigol05, minguzzi05, wilson_20}, those that result in (generalized) thermalization exhibit a rapid growth of entanglement; their entanglement entropy grows linearly in time~\cite{dechiara2006entanglement, fagotti2008evolution, lauchli2008spreading, kim2013ballistic, alba_calabrese_17}. Since the evolution time $t$ after the quench enters the emergent Hamiltonian as a parameter, this means that the desired eigenstate must have an entanglement entropy that is proportional to a Hamiltonian parameter. 

To introduce the emergent local Hamiltonians, let us consider a quantum quench (at $t=0$) from Hamiltonian $\hat H^{(0)}\rightarrow\hat H^f$ [both local, namely, they are extensive sums of operators with support on ${\cal O}(1)$ sites], for an initial state $|\psi_0 \rangle$ that is an eigenstate of $\hat H^{(0)}$. At $t>0$, $|\psi(t)\rangle=\exp(-i\hat H^f t)|\psi_0\rangle$ is an eigenstate of $\hat{\cal M}(t)= \exp(-i \hat H^f t)\hat H^{(0)} \exp(i \hat H^f t)$ (we set $\hbar \equiv 1$). $\hat{\cal M}(t)$ is in general  highly nonlocal, and can be written as
\begin{equation} \label{def_M}
\hat{\cal M}(t) = \hat H^{(0)} + \sum_{n=1}^{\infty}\frac{(-it)^n}{n!}\hat {\cal H}_n \,,
\end{equation}
where $\hat {\cal H}_n=[\hat H^f,[\hat H^f,\cdots,[\hat H^f,\hat H^{(0)}]_{\cdots}]]$ is a nested $n$-order commutator. If $\hat {\cal H}_n$ vanishes for $n={\cal O}(1)$ then $\hat{\cal M}(t)$ is a local operator (as per our definition above). We then call $\hat{\cal H}(t)\equiv \hat{\cal M}(t)$ the emergent local Hamiltonian, of which $|\psi(t)\rangle$ is an eigenstate~\cite{vidmar_iyer_17}. In the context of geometric quenches, in which confining potentials are turned off, this description has allowed to understand and characterize low-entanglement far-from-equilibrium states~\cite{vidmar_iyer_17, vidmar_xu_17, zhang_vidmar_19}, and emergent Hamiltonians can be used to speed up quasi-adiabatic transformations~\cite{modak_vidmar_17}.

\begin{figure}[!t]
\includegraphics[width=0.98\columnwidth]{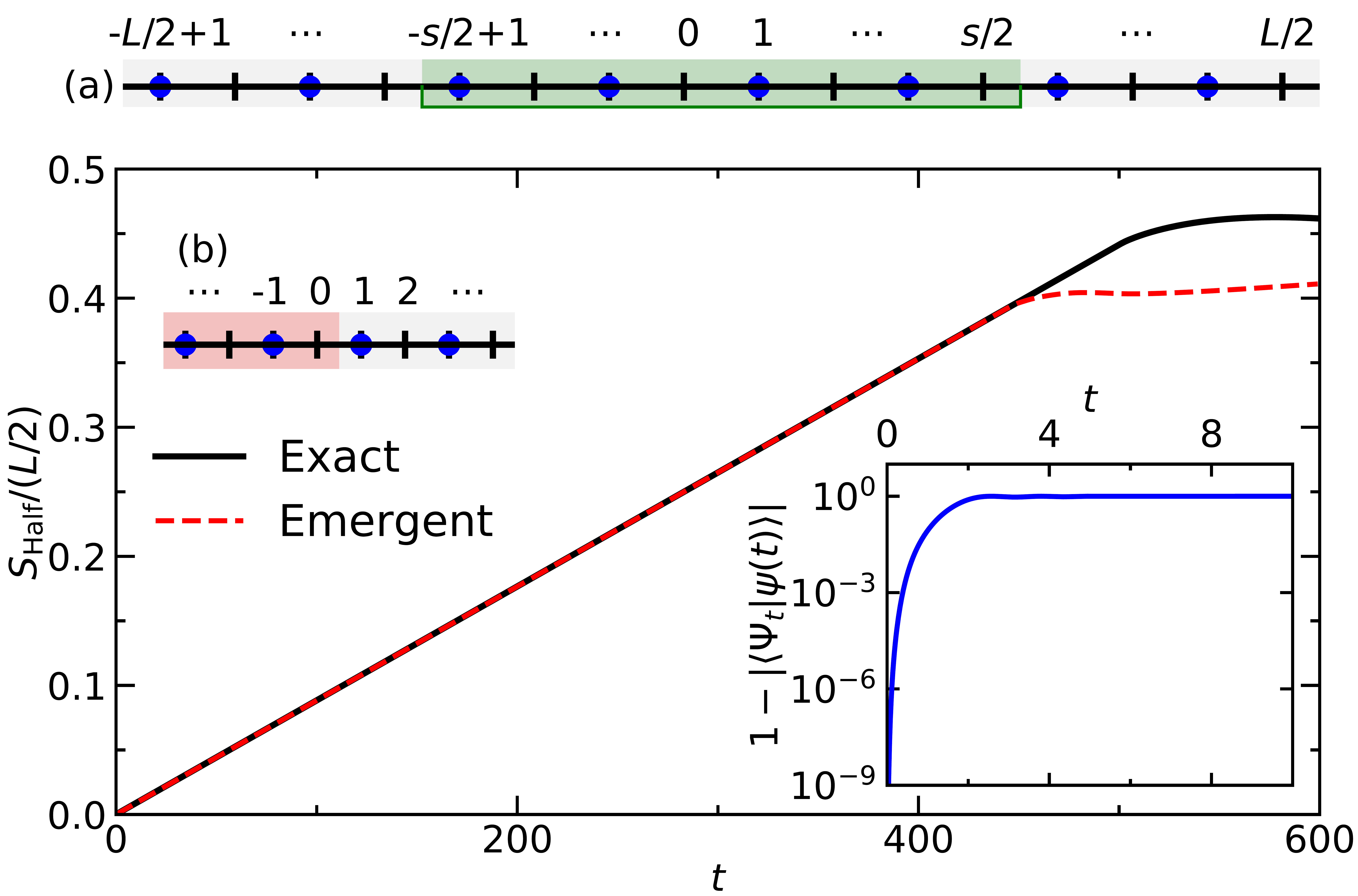}
\vspace{-0.1cm}
\caption{(a) Sketch of our initial density-wave state, which is a product state with odd sites occupied (blue dots) and even sites empty, in a 1D lattice with $L$ sites and open boundary conditions. The shaded area marks the central region of $s$ lattice sites in which most observables have their support. (b) Time evolution of the half-system von Neumann entanglement entropy $S_{\rm Half}(t)$. The lattice bipartition is sketched in the inset. We calculate $S_{\rm Half}(t)$ in the exact time-evolving state $|\psi(t)\rangle$ (solid line), and in the corresponding eigenstate $|\Psi_t\rangle$ of the emergent Hamiltonian $\hat{\cal H}(t)$ (dashed line), for $L=2000$. Inset: Subtracted overlap $1-|\langle \Psi_t | \psi(t) \rangle|$ vs $t$ for the states involved in the $S_{\rm Half}(t)$ calculations in the main panel.}\label{fig1}
\end{figure}

Our interest here are global quenches that produce highly entangled far-from-equilibrium states. We focus on 1D lattice systems with open boundary conditions, as described by the hard-core boson Hamiltonian:
\begin{equation}\label{Hhcb}
\hat{H} = -J\sum_{l=-L/2+1}^{L/2-1}{(\hat b^\dagger_l \hat b^{}_{l+1} + {\rm H.c.})}+ V \sum_{l=-L/2+1}^{L/2} (-1)^l b^\dagger_l \hat b^{}_{l}\,,
\end{equation}
where $\hat b^\dagger_l$ ($\hat b_l$) are the boson creation (annihilation) operators at site $l$ (they satisfy $\hat b^\dagger_l\hat b^\dagger_l=\hat b_l\hat b_l=0$), $J$ ($V$) is the hopping (local alternating potential) strength, and $L$ (even) is the number of lattice sites. We set $J\equiv1$, and the lattice spacing $a\equiv1$, in what follows. The initial state $|\psi_0\rangle$ for our quenches is taken to be the ground state of $\hat{H}^i=\hat{H}(V\gg 1)$ [odd (even) sites occupied (empty), see Fig.~\ref{fig1}(a)] and the time evolution is studied under $\hat{H}^f=\hat{H}(V=0)$. Dynamics of density-wave states like the one considered here have been studied in experiments with ultracold quantum gases in optical lattices~\cite{trotzky_chen_12, schreiber_hodgman_15, bordia_lueschen_16}, as well as theoretically~\cite{rigol_muramatsu_06, cramer_dawson_08, flesch_cramer_08}. We note that $|\psi_0\rangle$ for $V \to \infty$ is a highly excited eigenstate of 
\begin{equation}\label{H0hcb}
\hat{H}^{(0)} = \sum_{l=-L/2+1}^{L/2}{l \, \hat b^\dagger_l\hat b_l}\,,
\end{equation}
so our quench $\hat{H}^i\rightarrow\hat{H}^f$ starting from the ground state of $\hat{H}^i$ is equivalent to the quench $\hat{H}^{(0)}\rightarrow\hat{H}^f$ starting from a highly excited eigenstate of $\hat{H}^{(0)}$.

Mapping hard-core bosons onto spinless fermions $\hat b_l =e^{i\pi\sum_{m<l}\hat c_m^\dagger \hat c_m}\hat c_l$~\cite{cazalilla_citro_review_11}, one can reformulate this problem in the language of spinless fermions with initial Hamiltonian $\hat{H}_{\rm SF}^{(0)} = \sum_{l}{l \, \hat c^\dagger_l\hat c^{}_l}$ and final Hamiltonian $\hat{H}^f_{\rm SF}=-\sum_{l}{(\hat c^\dagger_l \hat c^{}_{l+1}+{\rm H.c.})}$. For those two Hamiltonians one can show that $\hat{\cal M}(t)=\hat{\cal H}(t)+\hat{B}(t)$, with an emergent Hamiltonian of the form~\cite{vidmar_iyer_17}
\begin{equation}\label{EH}
\hat{\cal H}(t) = \sum_{l=-L/2+1}^{L/2} l \, \hat c^\dagger_l\hat c^{}_l - t \sum_{l=-L/2+1}^{L/2-1}(i\hat c^\dagger_{l+1} \hat c^{}_{l} + {\rm H.c.}),
\end{equation}
in which the second term is the product of the time $t$ after the quench and the particle-current operator, and $\hat{B}(t) = \sum_{n=2}^{\infty}\frac{(-it)^n}{n!}\hat {\cal B}_n$. The operators $\hat {\cal B}_n$ are nonlocal one-particle operators, whose support grows from the open boundaries of our chains linearly with $n$~\cite{vidmar_iyer_17,suppmat}.

We define $|\Psi_t\rangle$ to be the many-body eigenstate of $\hat{\cal H}(t)$ that corresponds to $|\psi_0\rangle$ at $t=0$, which we can track as $t$ changes because we know its eigenenergy and its occupied single-particle eigenstates. The corrections to the overlap $|\langle \Psi_t | \psi(t) \rangle|$ that occur because of the ``boundary operator'' $\hat{B}(t)$ were shown to be exponentially small (in $L$) for extensive times (in $L$) for initial states of interest to geometric quenches~\cite{vidmar_iyer_17}. This is not the case for our initial density-wave state, and for other initial states of interest to the question of (generalized) thermalization, so one may think that $|\Psi_t\rangle$ is of no use for problems involving large entanglement production. Remarkably, because of the finite speed of propagation of information from the boundaries, $|\Psi_t\rangle$ provides exact predictions for the expectation values of observables in the bulk of large systems. With that in mind, we mostly focus on observables that have support on the central $s$ sites of the lattice, namely, with $j\in j_S = \{ -s/2+1, ..., s/2 \}$, as indicated by the shaded region in Fig.~\ref{fig1}(a).

\textit{Quantum dynamics.}
We first study the von Neumann entanglement entropy $S_{\rm Half}(t)$ for a bipartition of the system into two halves, see Fig.~\ref{fig1}(b). We calculate $S_{\rm Half}(t)$ numerically using the one-body correlation matrix of the half system~\cite{peschel_eisler_09} in the exact time-evolving state $|\psi(t)\rangle$ and in the corresponding eigenstate $|\Psi_t\rangle$ of the emergent Hamiltonian $\hat{\cal H}(t)$. The results in Fig.~\ref{fig1}(b) show that both entanglement entropies agree in the linear regime $S_{\rm Half}(t) \propto t$. They depart from each other at times close to $t^* \approx L/(2 v_{\rm max})$, where $v_{\rm max} = 2$ is the maximal group velocity, which is the time at which the entanglement entropy is expected to saturate to its extensive ($L$-dependent) value~\cite{calabrese_2005}. This is the time it takes for a particle with $v_{\rm max}$ to move from the boundaries to the center of the chain. In stark contrast (and as expected), the inset in Fig.~\ref{fig1}(b) shows that the overlap $|\langle \Psi_t | \psi(t) \rangle|$ vanishes in times that are ${\mathcal O}(1)$.

Next, we focus on observables for the hard-core bosons that are accessible in experiments with ultracold gases in optical lattices~\cite{bloch08, bakr_gillen_09, sherson_weitenberg_10}, namely, the site occupations and the quasimomentum distributions. While dynamics of the site occupations are identical for hard-core bosons and noninteracting fermions, dynamics of quasimomentum distributions are not. We study the integrated relative differences between observables in the exact time-evolving state $|\psi(t)\rangle$ and in the eigenstate $|\Psi_t \rangle$ of the emergent Hamiltonian ${\cal H}(t)$, defined as
\begin{equation} \label{def_dn}
\Delta n(s, t) = \frac{\sum_{j \in j_S} |n_j(t) - \tilde n_j(t)|}{\sum_{j \in j_S} n_j(t)} \,
\end{equation}
for site occupations, where $n_j(t) = \langle \psi(t) | \hat b_j^\dagger \hat b^{}_j |\psi(t) \rangle$ and $\tilde n_{j}(t) = \langle \Psi_t |  \hat b_j^\dagger \hat b_j^{}  | \Psi_t \rangle$, and
\begin{equation} \label{def_dm}
\Delta m(s, t) = \frac{\sum_{k \in k_S} |m_k(t) - \tilde m_k(t)|}{\sum_{k \in k_S} m_k(t)} \,
\end{equation}
for the quasimomentum distributions, defined as
\begin{equation}\label{mk}
m_k(t)=\frac{1}{s}\sum_{\{j,l\}\in j_S}{e^{ik(j-l)}C_{jl}(t)} \,
\end{equation}
for $|\psi(t) \rangle$, namely, as the Fourier transform of the one-body correlation matrix $C_{jl}(t)=\langle \psi(t) | \hat b^\dagger_j \hat b^{}_l | \psi(t) \rangle$ within the region with $s$ sites sketched in Fig.~\ref{fig1}(a). The number of sites $s$ is used to determine the set of $k_S$ numbers so that $\sum_{k \in k_S} m_k(t)$ is the number of bosons in that region. $\tilde m_k(t)$ in Eq.~(\ref{def_dm}) is the corresponding momentum distribution for $| \Psi_t \rangle$, obtained replacing $C_{jl}(t) \to \tilde C_{jl}(t)=\langle \Psi_t | \hat b^\dagger_j \hat b^{}_{l} | \Psi_t \rangle$ in Eq.~(\ref{mk}). All expectation values are calculated following the approach introduced in Refs.~\cite{rigol04, rigol_muramatsu_04sept}.

\begin{figure}[!t]
\includegraphics[width=0.98\columnwidth]{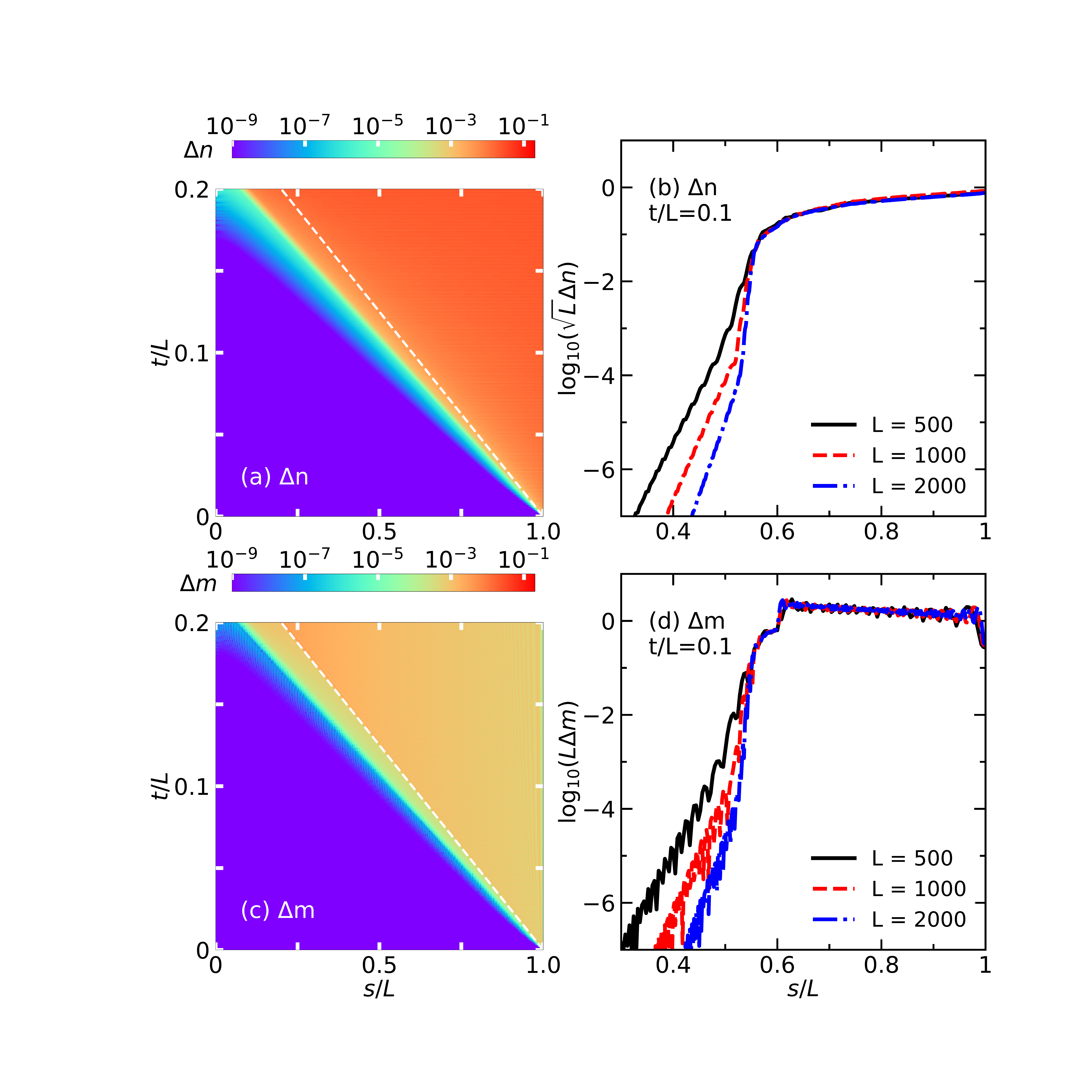}
\vspace{-0.2cm}
\caption{Integrated relative differences between observables in the exact time-evolving state $|\psi(t)\rangle$ and in the eigenstate $|\Psi_t \rangle$ of the emergent Hamiltonian. (a) and (c): Density plots (in log$_{10}$ scale) of $\Delta n(s,t)$ and $\Delta m(s,t)$ from Eqs.~(\ref{def_dn}) and~(\ref{def_dm}), respectively, for $L=2000$. White dashed lines depict $t/t^*= 1 - s/L$, where $t^* = L/4$. (b) and (d): Logarithm (base 10) of the scaled differences $\sqrt{L} \Delta n(s)$ and $L \Delta m(s)$, respectively, as functions of $s/L$, at a fixed time $t/L=0.1$ and for three different system sizes $L=500, 1000, 2000$.} \label{fig2}
\end{figure}

Figures~\ref{fig2}(a) and~\ref{fig2}(c) show density plots (notice the log$_{10}$ scale) of $\Delta n(s,t)$ and $\Delta m(s,t)$, respectively, as functions of the scaled volume $s/L$ and time $t/L$, for a lattice with $L=2000$ sites. As expected from the previous discussions, because of the boundary operator $\hat{B}(t)$, the region of applicability of the emergent eigenstate solution shrinks with increasing the scaled time $t/L$. Using the maximal group velocity $v_{\rm max} = 2$ allows one to estimate the region in which the emergent eigenstate solution is accurate (see the dashed line).

\begin{figure}[!t]
\begin{center}
\includegraphics[width=0.98\columnwidth]{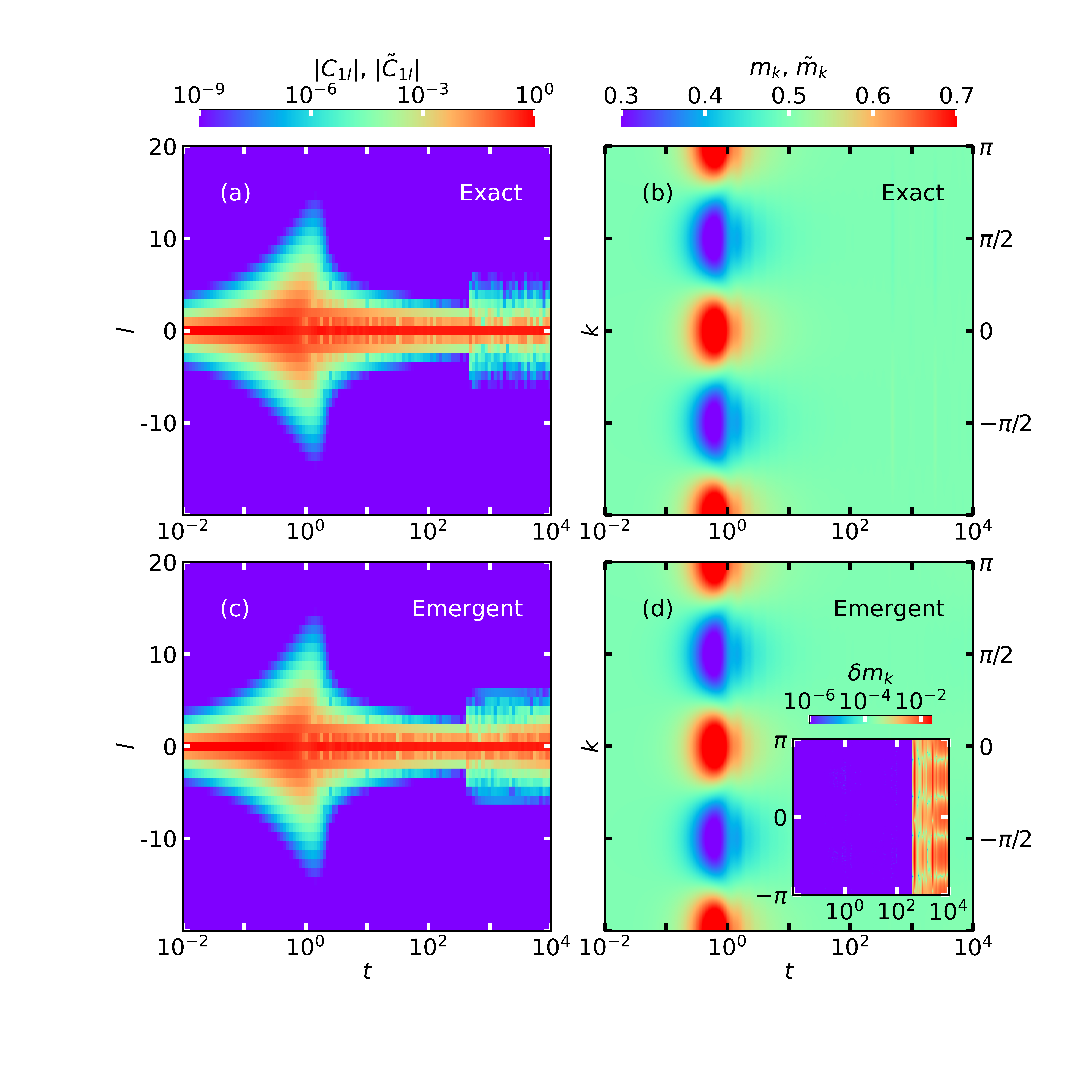}
\vspace{-0.1cm}
\caption{Dynamics of one-body correlations and momentum distributions in a lattice with $L=2000$ sites. (a) and (c), One-body correlations $|C_{1l}(t)| = |\langle \psi(t) | \hat b^\dagger_1 \hat b^{}_{l} | \psi(t) \rangle|$ and $|\tilde C_{1l}(t)| = |\langle \Psi_t | \hat b^\dagger_1 \hat b^{}_{l} | \Psi_t \rangle|$, respectively (measured with respect to site $j=1$). Due to the short-range nature of correlations, we only show results for $l \in [-20,20]$. (b) and (d) Quasimomentum distributions $m_k(t)$ and $\tilde m_k(t)$, respectively, for $s=100$. The inset in (d) shows the quasimomentum resolved difference $\delta m_k(t) = |m_k(t) - \tilde m_k(t)|/ m_k(t)$.}\label{fig3}
\end{center}
\end{figure}

The effect of increasing the system size $L$ at a fixed scaled time $t/L = 0.1$ is shown in Figs.~\ref{fig2}(b) and~\ref{fig2}(d). One can see that the scaled differences $\sqrt{L} \Delta n(s)$ and $L \Delta m(s)$ exhibit sharp transitions from vanishingly small values at $s/L \lesssim 0.6$ to nearly constant values at $s/L > 0.6$. $s/L=0.6$ is expected from $t/t^*= 1 - s/L$ for $t/L=0.1$. Most notably, for $s/L \lesssim 0.6$ the scaled differences decay rapidly with increasing $L$. This suggests that, so long as $s/L\lesssim 1-t/t^*$ for large system sizes, the emergent eigenstate solution provides a numerically exact description of the time evolution of observables in the central region of the lattice with $s$ sites.

On their way to the long-time equilibrated values, the dynamics of hard-core boson observables for our initial density-wave state and final homogeneous Hamiltonian is nontrivial, and it is fully captured by the emergent eigenstate solution. In Figs.~\ref{fig3}(a) and~\ref{fig3}(c), we show the time evolution of one-body correlations $|C_{1l}(t)|$ and $|\tilde C_{1l}(t)|$. Our initial state being a product state has no correlations, but they develop in time and reach their maximal extent at time $t \approx 1$, after which they decrease monotonically with time. As a result of the emergence of one-body correlations at short times after the quench, peaks (at $k=0$ and $\pi$) and dips (at $k=\pm\pi/2$) develop in the corresponding hard-core boson momentum distributions, shown in Figs.~\ref{fig3}(b) and~\ref{fig3}(d) (this does not happen for the fermions onto which they are mapped), and they disappear at long times as the system equilibrates to the generalized Gibbs ensemble prediction. For finite systems, such as the one considered in Fig.~\ref{fig3}, a revival in $|C_{1l}(t)|$ and $|\tilde C_{1l}(t)|$ occurs at $t^* \approx L/4$, when the light-cones starting at the boundaries of our chain reach the lattice center. This is about the time at which the emergent eigenstate description breaks down. Beyond that time, one can see differences between $|C_{1l}(t)|$ [Fig.~\ref{fig3}(a)] and $|\tilde C_{1l}(t)|$ [Fig.~\ref{fig3}(c)], and between $m_k(t)$ and $\tilde m_k(t)$ [inset in Fig.~\ref{fig3}(d)].

\textit{Generalized thermalization.} Let us conclude showing that the emergent eigenstate solution allows one to accurately describe the approach of observables to their long-time equilibrated values. Since hard-core bosons in 1D are integrable, the proper statistical ensemble to describe observables after equilibration is the GGE~\cite{rigol_dunjko_07, vidmar16},
\begin{equation}\label{gge}
\hat \rho_{\rm GGE}=\frac{1}{Z_{\rm GGE}}e^{-\sum_k{\lambda_k \hat I_k}}\,,
\end{equation}
where $Z_{\rm GGE}=\Tr[\exp(-\sum_k{\lambda_k \hat I_k})]$ is the partition function, $\{\hat I_k\}$ are one-body conserved quantities given by the occupations of single-particle eigenstates of the fermionic $H^f_{\rm SF}$ [defined below Eq.~(\ref{H0hcb})], and $\{ \lambda_k\}$ are the Lagrange multipliers set by the initial state, $\lambda_k = \ln [(1- \langle\psi_0|\hat I_k|\psi_0\rangle)/\langle\psi_0|\hat I_k|\psi_0\rangle]$. The GGE is particularly simple for our initial state because $\langle\psi_0 |\hat I_k|\psi_0 \rangle=0.5$, so that $\lambda_k=0$ for all $k$. Hence, the GGE is equivalent to a grand-canonical ensemble at infinite temperature~\cite{rigol_fitzpatrick_11}.

\begin{figure}[!t]
\includegraphics[width=0.98\columnwidth]{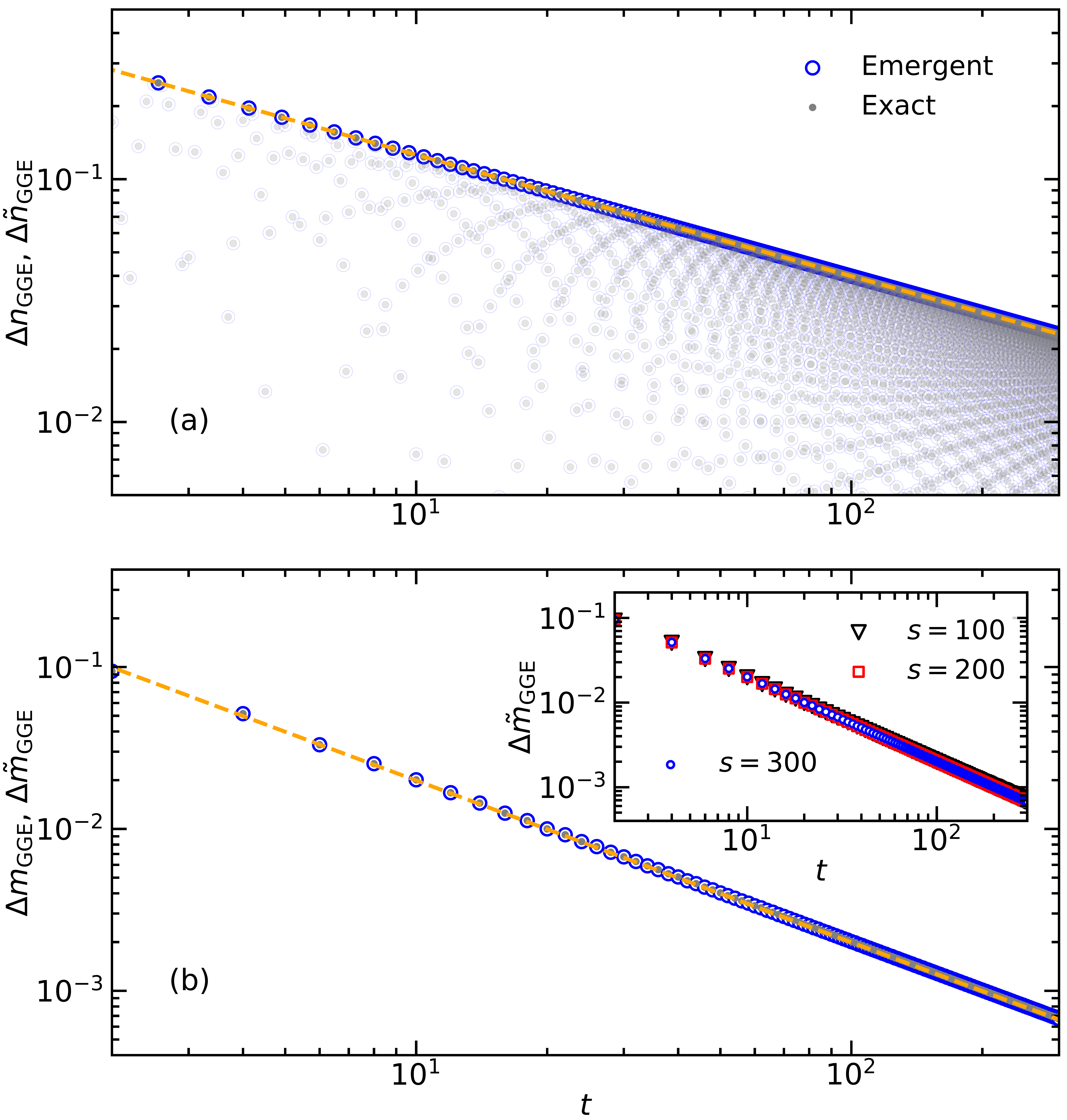}
\vspace{-0.1cm}
\caption{Generalized thermalization on a lattice with $L=2000$. (a) Shows $\Delta n_{\rm GGE}(s,t)=|J_0(4t)|$ and $\Delta \tilde n_{\rm GGE}(s,t)$, while (b) shows $\Delta m_{\rm GGE}(s,t)$ and $\Delta \tilde m_{\rm GGE}(s,t)$, vs $t$ for $s=300$. The dashed lines show the result of power-law fits $\propto t^{-\alpha}$ for $t \geq 100$, with $\alpha = 0.50$ in (a) and $\alpha = 1.00$ in (b). The inset in (b) shows $\Delta \tilde m_{\rm GGE}(s,t)$ for three values of $s$.}  \label{fig4}
\end{figure}

In analogy to Eqs.~(\ref{def_dn}) and~(\ref{def_dm}), we define the integrated relative difference of site occupations,
\begin{equation} \label{def_dn_gge}
\Delta n_{\rm GGE}(s, t) = \frac{ \sum_{j \in j_S} | n_j(t) - n^{\rm GGE}_j| } { \sum_{j \in j_S} n^{\rm GGE}_j}\,,
\end{equation}
and of the quasimomentum distribution,
\begin{equation} \label{def_dm_gge}
\Delta m_{\rm GGE}(s, t) = \frac{ \sum_{k \in k_S} | m_k(t) - m^{\rm GGE}_k | } { \sum_{k \in k_S}  m^{\rm GGE}_k }\,,
\end{equation}
where $n_j^{\rm GGE} = m_k^{\rm GGE} = 1/2$. Equivalently, we define $\Delta \tilde n_{\rm GGE}(s, t)$ by replacing $n_j(t)\rightarrow \tilde n_j(t)$ and $\Delta \tilde m_{\rm GGE}(s, t)$ by replacing $m_k(t) \rightarrow \tilde m_k(t)$. Since the site occupations of the hard-core bosons are identical to those of noninteracting fermions, it is straightforward to show that, in the thermodynamic limit, $\Delta n_{\rm GGE}(s,t)=|J_0(4t)|$, where $J_0$ is the zeroth-order Bessel function of the first kind. 

In Fig.~\ref{fig4}(a) [Fig.~\ref{fig4}(b)] we show results for  $\Delta n_{\rm GGE}(s,t)=|J_0(4t)|$ and $\Delta \tilde n_{\rm GGE}(s,t)$ [$\Delta m_{\rm GGE}(s,t)$ and $\Delta \tilde m_{\rm GGE}(s,t)$] vs $t$ for $s=300$ in a system with $L=2000$. The numerical results obtained within the emergent eigenstate solution are indistinguishable from the analytical results for $\Delta n_{\rm GGE}(s,t)$ and from the exact numerical results for $\Delta m_{\rm GGE}(s,t)$. The inset in Fig.~\ref{fig4}(b) shows that the finite-size effects associated to the value of $s$ selected are negligible in $\Delta \tilde m_{\rm GGE}(s,t)$. 

We note that the relative differences vanish as power laws in time $t^{-\alpha}$, with $\alpha = 0.50$ for $\Delta n_{\rm GGE}(s,t)$ and $\Delta \tilde n_{\rm GGE}(s,t)$, and $\alpha = 1.00$ for $\Delta m_{\rm GGE}(s,t)$ and $\Delta \tilde m_{\rm GGE}(s,t)$. Power-law approaches to the GGE predictions are common in integrable models like the one considered here~\cite{vidmar16, essler_fagotti_2016, gramsch_rigol_12, murthy_srednicki_19, gluza_19}. For an accurate description of such power-law relaxations in the context of the emergent eigenstate solution, one can simply increase $L$ keeping $t/L \leq {\rm const}$, such that $\Delta \tilde n_{\rm GGE}(s,t)$ and $\Delta \tilde m_{\rm GGE}(s,t)$ can be made arbitrary small while ensuring that the emergent eigenstate solution is numerically exact.

\textit{Summary.} We have shown that it is possible to construct emergent local Hamiltonians of which one eigenstate describes the entire generalized thermalization process following a quantum quench that produces extensive entanglement. Namely, the ``emergent'' eigenstate mirrors both the far from equilibrium stages as well as the equilibrated one. In the specific construction considered here, the dynamics in a subsystem of size $s$ is correctly described by the emergent eigenstate for times $t/t^*\lesssim 1 - s/L$ with $t^*=L/(2v_\text{max})$, where $v_\text{max}$ is the maximal group velocity in the lattice. A novel byproduct of our construction is that, for $t<t^*$, our eigenstate of interest exhibits an entanglement entropy that is linear with $t$, which is a parameter in the emergent Hamiltonian. This is to be contrasted to the traditional area- vs volume-law discussion about the entanglement entropy of Hamiltonian eigenstates. Here the entanglement entropy is a (linear) function of a Hamiltonian parameter.

The ultimate limits of emergent constructions and their applicability remain to be explored. In the context of geometric quenches, emergent eigenstate solutions have been provided for a wide range of trapped hard-core boson and spinless fermion systems in the ground state and at nonzero temperature~\cite{vidmar_iyer_17, vidmar_xu_17, modak_vidmar_17}, and for domain walls in spin-1/2 XXZ chains~\cite{vidmar_iyer_17}. Our results here open the door to using emergent eigenstate solutions to explore global quenches, which due to the high entanglement they produce are much more challenging to study computationally than geometric quenches.

\textit{Acknowledgments.}
This work was supported by the National Science Foundation under Grant No.~PHY-2012145 (Y.Z. and M.R.), and by the the Slovenian Research Agency (ARRS), Research core fundings Grants No.~P1-0044 and No.~J1-1696 (L.V.).


\bibliographystyle{biblev1}
\bibliography{references}

\begin{thebibliography}{10}
\expandafter\ifx\csname url\endcsname\relax
  \def\url#1{{\tt #1}}\fi
\expandafter\ifx\csname urlprefix\endcsname\relax\def\urlprefix{URL }\fi
\expandafter\ifx\csname bibinfo\endcsname\relax\def\bibinfo#1#2{#2}\fi
\expandafter\ifx\csname eprint\endcsname\relax\def\eprint#1{\url{#1}}\fi

\bibitem{polkovnikov_sengupta_review_11}
\bibinfo{author}{A.~Polkovnikov}, \bibinfo{author}{K.~Sengupta},
  \bibinfo{author}{A.~Silva}, and \bibinfo{author}{M.~Vengalattore},
  \bibinfo{title}{\textit{Colloquium}: Nonequilibrium dynamics of closed
  interacting quantum systems},
  \bibinfo{journal}{\href{http://dx.doi.org/10.1103/RevModPhys.83.863}{Rev.
  Mod. Phys.}} \href{http://dx.doi.org/10.1103/RevModPhys.83.863}{{\bf
  \bibinfo{volume}{83}}, \bibinfo{pages}{863}}
  (\href{http://dx.doi.org/10.1103/RevModPhys.83.863}{\bibinfo{year}{2011}}).

\bibitem{eisert_friesdorf_review_15}
\bibinfo{author}{J.~Eisert}, \bibinfo{author}{M.~Friesdorf}, and
  \bibinfo{author}{C.~Gogolin}, \bibinfo{title}{Quantum many-body systems out
  of equilibrium},
  \bibinfo{journal}{\href{http://dx.doi.org/10.1038/nphys3215}{Nature Physics}}
  \href{http://dx.doi.org/10.1038/nphys3215}{{\bf \bibinfo{volume}{11}},
  \bibinfo{pages}{124}}
  (\href{http://dx.doi.org/10.1038/nphys3215}{\bibinfo{year}{2015}}).

\bibitem{dalessio_kafri_16}
\bibinfo{author}{L.~D'Alessio}, \bibinfo{author}{Y.~Kafri},
  \bibinfo{author}{A.~Polkovnikov}, and \bibinfo{author}{M.~Rigol},
  \bibinfo{title}{From quantum chaos and eigenstate thermalization to
  statistical mechanics and thermodynamics},
  \bibinfo{journal}{\href{http://dx.doi.org/10.1080/00018732.2016.1198134}{Adv.
  Phys.}} \href{http://dx.doi.org/10.1080/00018732.2016.1198134}{{\bf
  \bibinfo{volume}{65}}, \bibinfo{pages}{239}}
  (\href{http://dx.doi.org/10.1080/00018732.2016.1198134}{\bibinfo{year}{2016}}).

\bibitem{mori_ikeda_18}
\bibinfo{author}{T.~Mori}, \bibinfo{author}{T.~N. Ikeda},
  \bibinfo{author}{E.~Kaminishi}, and \bibinfo{author}{M.~Ueda},
  \bibinfo{title}{Thermalization and prethermalization in isolated quantum
  systems: a theoretical overview},
  \bibinfo{journal}{\href{http://dx.doi.org/10.1088/1361-6455/aabcdf}{J. Phys.
  B}} \href{http://dx.doi.org/10.1088/1361-6455/aabcdf}{{\bf
  \bibinfo{volume}{51}}, \bibinfo{pages}{112001}}
  (\href{http://dx.doi.org/10.1088/1361-6455/aabcdf}{\bibinfo{year}{2018}}).

\bibitem{deutsch_18}
\bibinfo{author}{J.~M. Deutsch}, \bibinfo{title}{Eigenstate thermalization
  hypothesis},
  \bibinfo{journal}{\href{http://dx.doi.org/10.1088/1361-6633/aac9f1}{Rep.
  Prog. Phys.}} \href{http://dx.doi.org/10.1088/1361-6633/aac9f1}{{\bf
  \bibinfo{volume}{81}}, \bibinfo{pages}{082001}}
  (\href{http://dx.doi.org/10.1088/1361-6633/aac9f1}{\bibinfo{year}{2018}}).

\bibitem{kinoshita_wenger_06}
\bibinfo{author}{T.~Kinoshita}, \bibinfo{author}{T.~Wenger}, and
  \bibinfo{author}{D.~S. Weiss}, \bibinfo{title}{A quantum {Newton's} cradle},
  \bibinfo{journal}{\href{http://dx.doi.org/10.1038/nature04693}{Nature}}
  \href{http://dx.doi.org/10.1038/nature04693}{{\bf \bibinfo{volume}{440}},
  \bibinfo{pages}{900}}
  (\href{http://dx.doi.org/10.1038/nature04693}{\bibinfo{year}{2006}}).

\bibitem{gring_kuhnert_12}
\bibinfo{author}{M.~Gring}, \bibinfo{author}{M.~Kuhnert},
  \bibinfo{author}{T.~Langen}, \bibinfo{author}{T.~Kitagawa},
  \bibinfo{author}{B.~Rauer}, \bibinfo{author}{M.~Schreitl},
  \bibinfo{author}{I.~Mazets}, \bibinfo{author}{D.~A. Smith},
  \bibinfo{author}{E.~Demler}, and \bibinfo{author}{J.~Schmiedmayer},
  \bibinfo{title}{Relaxation and prethermalization in an isolated quantum
  system},
  \bibinfo{journal}{\href{http://dx.doi.org/10.1126/science.1224953}{Science}}
  \href{http://dx.doi.org/10.1126/science.1224953}{{\bf \bibinfo{volume}{337}},
  \bibinfo{pages}{1318}}
  (\href{http://dx.doi.org/10.1126/science.1224953}{\bibinfo{year}{2012}}).

\bibitem{trotzky_chen_12}
\bibinfo{author}{S.~Trotzky}, \bibinfo{author}{Y.-A. Chen},
  \bibinfo{author}{A.~Flesch}, \bibinfo{author}{I.~P. McCulloch},
  \bibinfo{author}{U.~Schollw\"ock}, \bibinfo{author}{J.~Eisert}, and
  \bibinfo{author}{I.~Bloch}, \bibinfo{title}{Probing the relaxation towards
  equilibrium in an isolated strongly correlated one-dimensional
  $\mathrm{B}$ose gas},
  \bibinfo{journal}{\href{http://dx.doi.org/10.1038/nphys2232}{Nat. Phys.}}
  \href{http://dx.doi.org/10.1038/nphys2232}{{\bf \bibinfo{volume}{8}},
  \bibinfo{pages}{325}}
  (\href{http://dx.doi.org/10.1038/nphys2232}{\bibinfo{year}{2012}}).

\bibitem{meinert13}
\bibinfo{author}{F.~Meinert}, \bibinfo{author}{M.~J. Mark},
  \bibinfo{author}{E.~Kirilov}, \bibinfo{author}{K.~Lauber},
  \bibinfo{author}{P.~Weinmann}, \bibinfo{author}{A.~J. Daley}, and
  \bibinfo{author}{H.-C. N\"agerl}, \bibinfo{title}{Quantum quench in an atomic
  one-dimensional {Ising} chain},
  \bibinfo{journal}{\href{http://dx.doi.org/10.1103/PhysRevLett.111.053003}{Phys.
  Rev. Lett.}} \href{http://dx.doi.org/10.1103/PhysRevLett.111.053003}{{\bf
  \bibinfo{volume}{111}}, \bibinfo{pages}{053003}}
  (\href{http://dx.doi.org/10.1103/PhysRevLett.111.053003}{\bibinfo{year}{2013}}).

\bibitem{langen_erne_15}
\bibinfo{author}{T.~Langen}, \bibinfo{author}{S.~Erne},
  \bibinfo{author}{R.~Geiger}, \bibinfo{author}{B.~Rauer},
  \bibinfo{author}{T.~Schweigler}, \bibinfo{author}{M.~Kuhnert},
  \bibinfo{author}{W.~Rohringer}, \bibinfo{author}{I.~E. Mazets},
  \bibinfo{author}{T.~Gasenzer}, and \bibinfo{author}{J.~Schmiedmayer},
  \bibinfo{title}{Experimental observation of a generalized {G}ibbs ensemble},
  \bibinfo{journal}{\href{http://dx.doi.org/10.1126/science.1257026}{Science}}
  \href{http://dx.doi.org/10.1126/science.1257026}{{\bf \bibinfo{volume}{348}},
  \bibinfo{pages}{207}}
  (\href{http://dx.doi.org/10.1126/science.1257026}{\bibinfo{year}{2015}}).

\bibitem{clos_porras_16}
\bibinfo{author}{G.~Clos}, \bibinfo{author}{D.~Porras},
  \bibinfo{author}{U.~Warring}, and \bibinfo{author}{T.~Schaetz},
  \bibinfo{title}{Time-resolved observation of thermalization in an isolated
  quantum system},
  \bibinfo{journal}{\href{http://dx.doi.org/10.1103/PhysRevLett.117.170401}{Phys.
  Rev. Lett.}} \href{http://dx.doi.org/10.1103/PhysRevLett.117.170401}{{\bf
  \bibinfo{volume}{117}}, \bibinfo{pages}{170401}}
  (\href{http://dx.doi.org/10.1103/PhysRevLett.117.170401}{\bibinfo{year}{2016}}).

\bibitem{kaufman_tai_16}
\bibinfo{author}{A.~M. Kaufman}, \bibinfo{author}{M.~E. Tai},
  \bibinfo{author}{A.~Lukin}, \bibinfo{author}{M.~Rispoli},
  \bibinfo{author}{R.~Schittko}, \bibinfo{author}{P.~M. Preiss}, and
  \bibinfo{author}{M.~Greiner}, \bibinfo{title}{Quantum thermalization through
  entanglement in an isolated many-body system},
  \bibinfo{journal}{\href{http://dx.doi.org/10.1126/science.aaf6725}{Science}}
  \href{http://dx.doi.org/10.1126/science.aaf6725}{{\bf \bibinfo{volume}{353}},
  \bibinfo{pages}{794}}
  (\href{http://dx.doi.org/10.1126/science.aaf6725}{\bibinfo{year}{2016}}).

\bibitem{Neill2016}
\bibinfo{author}{C.~Neill}, \bibinfo{author}{P.~Roushan},
  \bibinfo{author}{M.~Fang}, \bibinfo{author}{Y.~Chen},
  \bibinfo{author}{M.~Kolodrubetz}, \bibinfo{author}{Z.~Chen},
  \bibinfo{author}{A.~Megrant}, \bibinfo{author}{R.~Barends},
  \bibinfo{author}{B.~Campbell}, \bibinfo{author}{B.~Chiaro},
  \bibinfo{author}{A.~Dunsworth}, \bibinfo{author}{E.~Jeffrey},
  \bibinfo{author}{J.~Kelly}, \bibinfo{author}{J.~Mutus},
  \bibinfo{author}{P.~J.~J. O’Malley}, \bibinfo{author}{C.~Quintana},
  \bibinfo{author}{D.~Sank}, \bibinfo{author}{A.~Vainsencher},
  \bibinfo{author}{J.~Wenner}, \bibinfo{author}{T.~C. White},
  \bibinfo{author}{A.~Polkovnikov}, and \bibinfo{author}{J.~M. Martinis},
  \bibinfo{title}{Ergodic dynamics and thermalization in an isolated quantum
  system}, \bibinfo{journal}{\href{http://dx.doi.org/10.1038/nphys3830}{Nat.
  Phys.}} \href{http://dx.doi.org/10.1038/nphys3830}{{\bf
  \bibinfo{volume}{12}}, \bibinfo{pages}{1037}}
  (\href{http://dx.doi.org/10.1038/nphys3830}{\bibinfo{year}{2016}}).

\bibitem{tang_kao_18}
\bibinfo{author}{Y.~Tang}, \bibinfo{author}{W.~Kao}, \bibinfo{author}{K.-Y.
  Li}, \bibinfo{author}{S.~Seo}, \bibinfo{author}{K.~Mallayya},
  \bibinfo{author}{M.~Rigol}, \bibinfo{author}{S.~Gopalakrishnan}, and
  \bibinfo{author}{B.~L. Lev}, \bibinfo{title}{{Thermalization near
  Integrability in a Dipolar Quantum {Newton's} Cradle}},
  \bibinfo{journal}{\href{http://dx.doi.org/10.1103/PhysRevX.8.021030}{Phys.
  Rev. X}} \href{http://dx.doi.org/10.1103/PhysRevX.8.021030}{{\bf
  \bibinfo{volume}{8}}, \bibinfo{pages}{021030}}
  (\href{http://dx.doi.org/10.1103/PhysRevX.8.021030}{\bibinfo{year}{2018}}).

\bibitem{malvania2020}
\bibinfo{author}{N.~Malvania}, \bibinfo{author}{Y.~Zhang},
  \bibinfo{author}{Y.~Le}, \bibinfo{author}{J.~Dubail},
  \bibinfo{author}{M.~Rigol}, and \bibinfo{author}{D.~S. Weiss},
  \bibinfo{title}{Generalized hydrodynamics in strongly interacting {1D Bose}
  gases},
  \href{https://arxiv.org/abs/2009.06651}{\bibinfo{howpublished}{arXiv:2009.06651}}.

\bibitem{rigol_dunjko_07}
\bibinfo{author}{M.~Rigol}, \bibinfo{author}{V.~Dunjko},
  \bibinfo{author}{V.~Yurovsky}, and \bibinfo{author}{M.~Olshanii},
  \bibinfo{title}{Relaxation in a completely integrable many-body quantum
  system: An ab initio study of the dynamics of the highly excited states of
  {1D} lattice hard-core bosons},
  \bibinfo{journal}{\href{http://dx.doi.org/10.1103/PhysRevLett.98.050405}{Phys.
  Rev. Lett.}} \href{http://dx.doi.org/10.1103/PhysRevLett.98.050405}{{\bf
  \bibinfo{volume}{98}}, \bibinfo{pages}{050405}}
  (\href{http://dx.doi.org/10.1103/PhysRevLett.98.050405}{\bibinfo{year}{2007}}).

\bibitem{cazalilla_2006}
\bibinfo{author}{M.~A. Cazalilla}, \bibinfo{title}{Effect of suddenly turning
  on interactions in the luttinger model},
  \bibinfo{journal}{\href{http://dx.doi.org/10.1103/PhysRevLett.97.156403}{Phys.
  Rev. Lett.}} \href{http://dx.doi.org/10.1103/PhysRevLett.97.156403}{{\bf
  \bibinfo{volume}{97}}, \bibinfo{pages}{156403}}
  (\href{http://dx.doi.org/10.1103/PhysRevLett.97.156403}{\bibinfo{year}{2006}}).

\bibitem{calabrese_essler_2011}
\bibinfo{author}{P.~Calabrese}, \bibinfo{author}{F.~H.~L. Essler}, and
  \bibinfo{author}{M.~Fagotti}, \bibinfo{title}{Quantum quench in the
  transverse-field {I}sing chain},
  \bibinfo{journal}{\href{http://dx.doi.org/10.1103/PhysRevLett.106.227203}{Phys.
  Rev. Lett.}} \href{http://dx.doi.org/10.1103/PhysRevLett.106.227203}{{\bf
  \bibinfo{volume}{106}}, \bibinfo{pages}{227203}}
  (\href{http://dx.doi.org/10.1103/PhysRevLett.106.227203}{\bibinfo{year}{2011}}).

\bibitem{ilievski15}
\bibinfo{author}{E.~Ilievski}, \bibinfo{author}{J.~De~Nardis},
  \bibinfo{author}{B.~Wouters}, \bibinfo{author}{J.-S. Caux},
  \bibinfo{author}{F.~H.~L. Essler}, and \bibinfo{author}{T.~Prosen},
  \bibinfo{title}{Complete {generalized Gibbs} ensembles in an interacting
  theory},
  \bibinfo{journal}{\href{http://dx.doi.org/10.1103/PhysRevLett.115.157201}{Phys.
  Rev. Lett.}} \href{http://dx.doi.org/10.1103/PhysRevLett.115.157201}{{\bf
  \bibinfo{volume}{115}}, \bibinfo{pages}{157201}}
  (\href{http://dx.doi.org/10.1103/PhysRevLett.115.157201}{\bibinfo{year}{2015}}).

\bibitem{piroli_vernier_16}
\bibinfo{author}{L.~Piroli}, \bibinfo{author}{E.~Vernier}, and
  \bibinfo{author}{P.~Calabrese}, \bibinfo{title}{{Exact steady states for
  quantum quenches in integrable Heisenberg spin chains}},
  \bibinfo{journal}{\href{http://dx.doi.org/10.1103/PhysRevB.94.054313}{Phys.
  Rev. B}} \href{http://dx.doi.org/10.1103/PhysRevB.94.054313}{{\bf
  \bibinfo{volume}{94}}, \bibinfo{pages}{054313}}
  (\href{http://dx.doi.org/10.1103/PhysRevB.94.054313}{\bibinfo{year}{2016}}).

\bibitem{ilievski_quinn_17}
\bibinfo{author}{E.~Ilievski}, \bibinfo{author}{E.~Quinn}, and
  \bibinfo{author}{J.-S. Caux}, \bibinfo{title}{From interacting particles to
  equilibrium statistical ensembles},
  \bibinfo{journal}{\href{http://dx.doi.org/10.1103/PhysRevB.95.115128}{Phys.
  Rev. B}} \href{http://dx.doi.org/10.1103/PhysRevB.95.115128}{{\bf
  \bibinfo{volume}{95}}, \bibinfo{pages}{115128}}
  (\href{http://dx.doi.org/10.1103/PhysRevB.95.115128}{\bibinfo{year}{2017}}).

\bibitem{piroli_vernier_17}
\bibinfo{author}{L.~Piroli}, \bibinfo{author}{E.~Vernier},
  \bibinfo{author}{P.~Calabrese}, and \bibinfo{author}{M.~Rigol},
  \bibinfo{title}{{Correlations and diagonal entropy after quantum quenches in
  XXZ chains}},
  \bibinfo{journal}{\href{http://dx.doi.org/10.1103/PhysRevB.95.054308}{Phys.
  Rev. B}} \href{http://dx.doi.org/10.1103/PhysRevB.95.054308}{{\bf
  \bibinfo{volume}{95}}, \bibinfo{pages}{054308}}
  (\href{http://dx.doi.org/10.1103/PhysRevB.95.054308}{\bibinfo{year}{2017}}).

\bibitem{pozsgay_vernier_17}
\bibinfo{author}{B.~Pozsgay}, \bibinfo{author}{E.~Vernier}, and
  \bibinfo{author}{M.~A. Werner}, \bibinfo{title}{{On generalized Gibbs
  ensembles with an infinite set of conserved charges}},
  \bibinfo{journal}{\href{http://dx.doi.org/10.1088/1742-5468/aa82c1}{J. Stat.
  Mech.}} \href{http://dx.doi.org/10.1088/1742-5468/aa82c1}{{\bf
  \bibinfo{volume}{{\rm (2017)}}}, \bibinfo{pages}{093103}}.

\bibitem{deutsch_91}
\bibinfo{author}{J.~M. Deutsch}, \bibinfo{title}{Quantum statistical mechanics
  in a closed system},
  \bibinfo{journal}{\href{http://dx.doi.org/10.1103/PhysRevA.43.2046}{Phys.
  Rev. A}} \href{http://dx.doi.org/10.1103/PhysRevA.43.2046}{{\bf
  \bibinfo{volume}{43}}, \bibinfo{pages}{2046}}
  (\href{http://dx.doi.org/10.1103/PhysRevA.43.2046}{\bibinfo{year}{1991}}).

\bibitem{srednicki_94}
\bibinfo{author}{M.~Srednicki}, \bibinfo{title}{Chaos and quantum
  thermalization},
  \bibinfo{journal}{\href{http://dx.doi.org/10.1103/PhysRevE.50.888}{Phys. Rev.
  E}} \href{http://dx.doi.org/10.1103/PhysRevE.50.888}{{\bf
  \bibinfo{volume}{50}}, \bibinfo{pages}{888}}
  (\href{http://dx.doi.org/10.1103/PhysRevE.50.888}{\bibinfo{year}{1994}}).

\bibitem{rigol_dunjko_08}
\bibinfo{author}{M.~Rigol}, \bibinfo{author}{V.~Dunjko}, and
  \bibinfo{author}{M.~Olshanii}, \bibinfo{title}{Thermalization and its
  mechanism for generic isolated quantum systems},
  \bibinfo{journal}{\href{http://dx.doi.org/10.1038/nature06838}{Nature}}
  \href{http://dx.doi.org/10.1038/nature06838}{{\bf \bibinfo{volume}{452}},
  \bibinfo{pages}{854}}
  (\href{http://dx.doi.org/10.1038/nature06838}{\bibinfo{year}{2008}}).

\bibitem{cassidy_clark_11}
\bibinfo{author}{A.~C. Cassidy}, \bibinfo{author}{C.~W. Clark}, and
  \bibinfo{author}{M.~Rigol}, \bibinfo{title}{Generalized thermalization in an
  integrable lattice system},
  \bibinfo{journal}{\href{http://dx.doi.org/10.1103/PhysRevLett.106.140405}{Phys.
  Rev. Lett.}} \href{http://dx.doi.org/10.1103/PhysRevLett.106.140405}{{\bf
  \bibinfo{volume}{106}}, \bibinfo{pages}{140405}}
  (\href{http://dx.doi.org/10.1103/PhysRevLett.106.140405}{\bibinfo{year}{2011}}).

\bibitem{caux_essler_13}
\bibinfo{author}{J.-S. Caux} and \bibinfo{author}{F.~H.~L. Essler},
  \bibinfo{title}{Time evolution of local observables after quenching to an
  integrable model},
  \bibinfo{journal}{\href{http://dx.doi.org/10.1103/PhysRevLett.110.257203}{Phys.
  Rev. Lett.}} \href{http://dx.doi.org/10.1103/PhysRevLett.110.257203}{{\bf
  \bibinfo{volume}{110}}, \bibinfo{pages}{257203}}
  (\href{http://dx.doi.org/10.1103/PhysRevLett.110.257203}{\bibinfo{year}{2013}}).

\bibitem{vidmar16}
\bibinfo{author}{L.~Vidmar} and \bibinfo{author}{M.~Rigol},
  \bibinfo{title}{{Generalized Gibbs ensemble in integrable lattice models}},
  \bibinfo{journal}{\href{http://dx.doi.org/10.1088/1742-5468/2016/06/064007}{J.
  Stat. Mech.}} \href{http://dx.doi.org/10.1088/1742-5468/2016/06/064007}{{\bf
  \bibinfo{volume}{{\rm (2016)}}}, \bibinfo{pages}{064007}}.

\bibitem{bukov_dalessio_review_15}
\bibinfo{author}{M.~Bukov}, \bibinfo{author}{L.~D'Alessio}, and
  \bibinfo{author}{A.~Polkovnikov}, \bibinfo{title}{Universal high-frequency
  behavior of periodically driven systems: from dynamical stabilization to
  floquet engineering},
  \bibinfo{journal}{\href{http://dx.doi.org/10.1080/00018732.2015.1055918}{Advances
  in Physics}} \href{http://dx.doi.org/10.1080/00018732.2015.1055918}{{\bf
  \bibinfo{volume}{64}}, \bibinfo{pages}{139}}
  (\href{http://dx.doi.org/10.1080/00018732.2015.1055918}{\bibinfo{year}{2015}}).

\bibitem{Oka2009}
\bibinfo{author}{T.~Oka} and \bibinfo{author}{H.~Aoki},
  \bibinfo{title}{{Photovoltaic Hall effect in graphene}},
  \bibinfo{journal}{\href{http://dx.doi.org/10.1103/PhysRevB.79.081406}{Phys.
  Rev. B}} \href{http://dx.doi.org/10.1103/PhysRevB.79.081406}{{\bf
  \bibinfo{volume}{79}}, \bibinfo{pages}{081406}}
  (\href{http://dx.doi.org/10.1103/PhysRevB.79.081406}{\bibinfo{year}{2009}}).

\bibitem{Kitagawa2010}
\bibinfo{author}{T.~Kitagawa}, \bibinfo{author}{E.~Berg},
  \bibinfo{author}{M.~Rudner}, and \bibinfo{author}{E.~Demler},
  \bibinfo{title}{{Topological characterization of periodically driven quantum
  systems}},
  \bibinfo{journal}{\href{http://dx.doi.org/10.1103/PhysRevB.82.235114}{Phys.
  Rev. B}} \href{http://dx.doi.org/10.1103/PhysRevB.82.235114}{{\bf
  \bibinfo{volume}{82}}, \bibinfo{pages}{235114}}
  (\href{http://dx.doi.org/10.1103/PhysRevB.82.235114}{\bibinfo{year}{2010}}).

\bibitem{Lindner2011}
\bibinfo{author}{N.~H. Lindner}, \bibinfo{author}{G.~Refael}, and
  \bibinfo{author}{V.~Galitski}, \bibinfo{title}{{Floquet topological insulator
  in semiconductor quantum wells}},
  \bibinfo{journal}{\href{http://dx.doi.org/10.1038/nphys1926}{Nature Phys.}}
  \href{http://dx.doi.org/10.1038/nphys1926}{{\bf \bibinfo{volume}{7}},
  \bibinfo{pages}{490}}
  (\href{http://dx.doi.org/10.1038/nphys1926}{\bibinfo{year}{2011}}).

\bibitem{Rechtsman2013}
\bibinfo{author}{M.~C. Rechtsman}, \bibinfo{author}{J.~M. Zeuner},
  \bibinfo{author}{Y.~Plotnik}, \bibinfo{author}{Y.~Lumer},
  \bibinfo{author}{D.~Podolsky}, \bibinfo{author}{F.~Dreisow},
  \bibinfo{author}{S.~Nolte}, \bibinfo{author}{M.~Segev}, and
  \bibinfo{author}{A.~Szameit}, \bibinfo{title}{{Photonic Floquet topological
  insulators}},
  \bibinfo{journal}{\href{http://dx.doi.org/10.1038/nature12066}{Nature}}
  \href{http://dx.doi.org/10.1038/nature12066}{{\bf \bibinfo{volume}{496}},
  \bibinfo{pages}{196}}
  (\href{http://dx.doi.org/10.1038/nature12066}{\bibinfo{year}{2013}}).

\bibitem{Jotzu2014}
\bibinfo{author}{G.~Jotzu}, \bibinfo{author}{M.~Messer},
  \bibinfo{author}{R.~Desbuquois}, \bibinfo{author}{M.~Lebrat},
  \bibinfo{author}{T.~Uehlinger}, \bibinfo{author}{D.~Greif}, and
  \bibinfo{author}{T.~Esslinger}, \bibinfo{title}{{Experimental realization of
  the topological Haldane model with ultracold fermions}},
  \bibinfo{journal}{\href{http://dx.doi.org/10.1038/nature13915}{Nature}}
  \href{http://dx.doi.org/10.1038/nature13915}{{\bf \bibinfo{volume}{515}},
  \bibinfo{pages}{237}}
  (\href{http://dx.doi.org/10.1038/nature13915}{\bibinfo{year}{2014}}).

\bibitem{dalessio_rigol_15}
\bibinfo{author}{L.~D'Alessio} and \bibinfo{author}{M.~Rigol},
  \bibinfo{title}{Dynamical preparation of floquet chern insulators},
  \bibinfo{journal}{\href{http://dx.doi.org/10.1038/ncomms9336}{Nat. Commun.}}
  \href{http://dx.doi.org/10.1038/ncomms9336}{{\bf \bibinfo{volume}{6}},
  \bibinfo{pages}{8336}}
  (\href{http://dx.doi.org/10.1038/ncomms9336}{\bibinfo{year}{2015}}).

\bibitem{rigol04}
\bibinfo{author}{M.~Rigol} and \bibinfo{author}{A.~Muramatsu},
  \bibinfo{title}{Emergence of quasicondensates of hard-core bosons at finite
  momentum},
  \bibinfo{journal}{\href{http://dx.doi.org/10.1103/PhysRevLett.93.230404}{Phys.
  Rev. Lett.}} \href{http://dx.doi.org/10.1103/PhysRevLett.93.230404}{{\bf
  \bibinfo{volume}{93}}, \bibinfo{pages}{230404}}
  (\href{http://dx.doi.org/10.1103/PhysRevLett.93.230404}{\bibinfo{year}{2004}}).

\bibitem{vidmar15}
\bibinfo{author}{L.~Vidmar}, \bibinfo{author}{J.~P. Ronzheimer},
  \bibinfo{author}{M.~Schreiber}, \bibinfo{author}{S.~Braun},
  \bibinfo{author}{S.~S. Hodgman}, \bibinfo{author}{S.~Langer},
  \bibinfo{author}{F.~Heidrich-Meisner}, \bibinfo{author}{I.~Bloch}, and
  \bibinfo{author}{U.~Schneider}, \bibinfo{title}{Dynamical quasicondensation
  of hard-core bosons at finite momenta},
  \bibinfo{journal}{\href{http://dx.doi.org/10.1103/PhysRevLett.115.175301}{Phys.
  Rev. Lett.}} \href{http://dx.doi.org/10.1103/PhysRevLett.115.175301}{{\bf
  \bibinfo{volume}{115}}, \bibinfo{pages}{175301}}
  (\href{http://dx.doi.org/10.1103/PhysRevLett.115.175301}{\bibinfo{year}{2015}}).

\bibitem{rigol05}
\bibinfo{author}{M.~Rigol} and \bibinfo{author}{A.~Muramatsu},
  \bibinfo{title}{Fermionization in an expanding 1$\mathrm{D}$ gas of hard-core
  bosons},
  \bibinfo{journal}{\href{http://dx.doi.org/10.1103/PhysRevLett.94.240403}{Phys.
  Rev. Lett.}} \href{http://dx.doi.org/10.1103/PhysRevLett.94.240403}{{\bf
  \bibinfo{volume}{94}}, \bibinfo{pages}{240403}}
  (\href{http://dx.doi.org/10.1103/PhysRevLett.94.240403}{\bibinfo{year}{2005}}).

\bibitem{minguzzi05}
\bibinfo{author}{A.~Minguzzi} and \bibinfo{author}{D.~M. Gangardt},
  \bibinfo{title}{Exact coherent states of a harmonically confined
  $\mathrm{T}$onks-$\mathrm{G}$irardeau gas},
  \bibinfo{journal}{\href{http://dx.doi.org/10.1103/PhysRevLett.94.240404}{Phys.
  Rev. Lett.}} \href{http://dx.doi.org/10.1103/PhysRevLett.94.240404}{{\bf
  \bibinfo{volume}{94}}, \bibinfo{pages}{240404}}
  (\href{http://dx.doi.org/10.1103/PhysRevLett.94.240404}{\bibinfo{year}{2005}}).

\bibitem{wilson_20}
\bibinfo{author}{J.~M. Wilson}, \bibinfo{author}{N.~Malvania},
  \bibinfo{author}{Y.~Le}, \bibinfo{author}{Y.~Zhang},
  \bibinfo{author}{M.~Rigol}, and \bibinfo{author}{D.~S. Weiss},
  \bibinfo{title}{Observation of dynamical fermionization},
  \bibinfo{journal}{\href{http://dx.doi.org/10.1126/science.aaz0242}{Science}}
  \href{http://dx.doi.org/10.1126/science.aaz0242}{{\bf \bibinfo{volume}{367}},
  \bibinfo{pages}{1461}}
  (\href{http://dx.doi.org/10.1126/science.aaz0242}{\bibinfo{year}{2020}}).

\bibitem{vidmar_iyer_17}
\bibinfo{author}{L.~Vidmar}, \bibinfo{author}{D.~Iyer}, and
  \bibinfo{author}{M.~Rigol}, \bibinfo{title}{{Emergent Eigenstate Solution to
  Quantum Dynamics Far from Equilibrium}},
  \bibinfo{journal}{\href{http://dx.doi.org/10.1103/PhysRevX.7.021012}{Phys.
  Rev. X}} \href{http://dx.doi.org/10.1103/PhysRevX.7.021012}{{\bf
  \bibinfo{volume}{7}}, \bibinfo{pages}{021012}}
  (\href{http://dx.doi.org/10.1103/PhysRevX.7.021012}{\bibinfo{year}{2017}}).

\bibitem{vidmar_xu_17}
\bibinfo{author}{L.~Vidmar}, \bibinfo{author}{W.~Xu}, and
  \bibinfo{author}{M.~Rigol}, \bibinfo{title}{Emergent eigenstate solution and
  emergent {G}ibbs ensemble for expansion dynamics in optical lattices},
  \bibinfo{journal}{\href{http://dx.doi.org/10.1103/PhysRevA.96.013608}{Phys.
  Rev. A}} \href{http://dx.doi.org/10.1103/PhysRevA.96.013608}{{\bf
  \bibinfo{volume}{96}}, \bibinfo{pages}{013608}}
  (\href{http://dx.doi.org/10.1103/PhysRevA.96.013608}{\bibinfo{year}{2017}}).

\bibitem{dechiara2006entanglement}
\bibinfo{author}{G.~DeChiara}, \bibinfo{author}{S.~Montangero},
  \bibinfo{author}{P.~Calabrese}, and \bibinfo{author}{R.~Fazio},
  \bibinfo{title}{Entanglement entropy dynamics of {H}eisenberg chains},
  \bibinfo{journal}{\href{http://dx.doi.org/10.1088/1742-5468/2006/03/P03001}{J.
  Stat. Mech.}} \href{http://dx.doi.org/10.1088/1742-5468/2006/03/P03001}{{\bf
  \bibinfo{volume}{2006}}, \bibinfo{pages}{P03001}}.

\bibitem{fagotti2008evolution}
\bibinfo{author}{M.~Fagotti} and \bibinfo{author}{P.~Calabrese},
  \bibinfo{title}{Evolution of entanglement entropy following a quantum quench:
  {A}nalytic results for the {$XY$} chain in a transverse magnetic field},
  \bibinfo{journal}{\href{http://dx.doi.org/10.1103/PhysRevA.78.010306}{Phys.
  Rev. A}} \href{http://dx.doi.org/10.1103/PhysRevA.78.010306}{{\bf
  \bibinfo{volume}{78}}, \bibinfo{pages}{010306}}
  (\href{http://dx.doi.org/10.1103/PhysRevA.78.010306}{\bibinfo{year}{2008}}).

\bibitem{lauchli2008spreading}
\bibinfo{author}{A.~M. L{\"a}uchli} and \bibinfo{author}{C.~Kollath},
  \bibinfo{title}{Spreading of correlations and entanglement after a quench in
  the one-dimensional {B}ose--{H}ubbard model},
  \bibinfo{journal}{\href{http://dx.doi.org/10.1088/1742-5468/2008/05/P05018}{J.
  Stat. Mech.}} \href{http://dx.doi.org/10.1088/1742-5468/2008/05/P05018}{{\bf
  \bibinfo{volume}{2008}}, \bibinfo{pages}{P05018}}
  (\href{http://dx.doi.org/10.1088/1742-5468/2008/05/P05018}{\bibinfo{year}{2008}}).

\bibitem{kim2013ballistic}
\bibinfo{author}{H.~Kim} and \bibinfo{author}{D.~A. Huse},
  \bibinfo{title}{Ballistic spreading of entanglement in a diffusive
  nonintegrable system},
  \bibinfo{journal}{\href{http://dx.doi.org/10.1103/PhysRevLett.111.127205}{Phys.
  Rev. Lett.}} \href{http://dx.doi.org/10.1103/PhysRevLett.111.127205}{{\bf
  \bibinfo{volume}{111}}, \bibinfo{pages}{127205}}
  (\href{http://dx.doi.org/10.1103/PhysRevLett.111.127205}{\bibinfo{year}{2013}}).

\bibitem{alba_calabrese_17}
\bibinfo{author}{V.~Alba} and \bibinfo{author}{P.~Calabrese},
  \bibinfo{title}{Entanglement and thermodynamics after a quantum quench in
  integrable systems},
  \href{http://dx.doi.org/10.1073/pnas.1703516114}{\bibinfo{journal}{\href{http://dx.doi.org/10.1073/pnas.1703516114}{Proc.
  Natl. Acad. Sci.}} }
  (\href{http://dx.doi.org/10.1073/pnas.1703516114}{\bibinfo{year}{2017}}).

\bibitem{zhang_vidmar_19}
\bibinfo{author}{Y.~Zhang}, \bibinfo{author}{L.~Vidmar}, and
  \bibinfo{author}{M.~Rigol}, \bibinfo{title}{Quantum dynamics of impenetrable
  {$\mathrm{SU}(N)$} fermions in one-dimensional lattices},
  \bibinfo{journal}{\href{http://dx.doi.org/10.1103/PhysRevA.99.063605}{Phys.
  Rev. A}} \href{http://dx.doi.org/10.1103/PhysRevA.99.063605}{{\bf
  \bibinfo{volume}{99}}, \bibinfo{pages}{063605}}
  (\href{http://dx.doi.org/10.1103/PhysRevA.99.063605}{\bibinfo{year}{2019}}).

\bibitem{modak_vidmar_17}
\bibinfo{author}{R.~Modak}, \bibinfo{author}{L.~Vidmar}, and
  \bibinfo{author}{M.~Rigol}, \bibinfo{title}{{Quantum adiabatic protocols
  using emergent local Hamiltonians}},
  \bibinfo{journal}{\href{http://dx.doi.org/10.1103/PhysRevE.96.042155}{Phys.
  Rev. E}} \href{http://dx.doi.org/10.1103/PhysRevE.96.042155}{{\bf
  \bibinfo{volume}{96}}, \bibinfo{pages}{042155}}
  (\href{http://dx.doi.org/10.1103/PhysRevE.96.042155}{\bibinfo{year}{2017}}).

\bibitem{schreiber_hodgman_15}
\bibinfo{author}{M.~Schreiber}, \bibinfo{author}{S.~S. Hodgman},
  \bibinfo{author}{P.~Bordia}, \bibinfo{author}{H.~P. L\"uschen},
  \bibinfo{author}{M.~H. Fischer}, \bibinfo{author}{R.~Vosk},
  \bibinfo{author}{E.~Altman}, \bibinfo{author}{U.~Schneider}, and
  \bibinfo{author}{I.~Bloch}, \bibinfo{title}{Observation of many-body
  localization of interacting fermions in a quasi-random optical lattice},
  \bibinfo{journal}{\href{http://dx.doi.org/10.1126/science.aaa7432}{Science}}
  \href{http://dx.doi.org/10.1126/science.aaa7432}{{\bf \bibinfo{volume}{349}},
  \bibinfo{pages}{842}}
  (\href{http://dx.doi.org/10.1126/science.aaa7432}{\bibinfo{year}{2015}}).

\bibitem{bordia_lueschen_16}
\bibinfo{author}{P.~Bordia}, \bibinfo{author}{H.~P. L\"uschen},
  \bibinfo{author}{S.~S. Hodgman}, \bibinfo{author}{M.~Schreiber},
  \bibinfo{author}{I.~Bloch}, and \bibinfo{author}{U.~Schneider},
  \bibinfo{title}{Coupling identical one-dimensional many-body localized
  systems},
  \bibinfo{journal}{\href{http://dx.doi.org/10.1103/PhysRevLett.116.140401}{Phys.
  Rev. Lett.}} \href{http://dx.doi.org/10.1103/PhysRevLett.116.140401}{{\bf
  \bibinfo{volume}{116}}, \bibinfo{pages}{140401}}
  (\href{http://dx.doi.org/10.1103/PhysRevLett.116.140401}{\bibinfo{year}{2016}}).

\bibitem{rigol_muramatsu_06}
\bibinfo{author}{M.~Rigol}, \bibinfo{author}{A.~Muramatsu}, and
  \bibinfo{author}{M.~Olshanii}, \bibinfo{title}{{Hard-core bosons on optical
  superlattices: Dynamics and relaxation in the superfluid and insulating
  regimes}},
  \bibinfo{journal}{\href{http://dx.doi.org/10.1103/PhysRevA.74.053616}{Phys.
  Rev. A}} \href{http://dx.doi.org/10.1103/PhysRevA.74.053616}{{\bf
  \bibinfo{volume}{74}}, \bibinfo{pages}{053616}}
  (\href{http://dx.doi.org/10.1103/PhysRevA.74.053616}{\bibinfo{year}{2006}}).

\bibitem{cramer_dawson_08}
\bibinfo{author}{M.~Cramer}, \bibinfo{author}{C.~M. Dawson},
  \bibinfo{author}{J.~Eisert}, and \bibinfo{author}{T.~J. Osborne},
  \bibinfo{title}{Exact relaxation in a class of nonequilibrium quantum lattice
  systems},
  \bibinfo{journal}{\href{http://dx.doi.org/10.1103/PhysRevLett.100.030602}{Phys.
  Rev. Lett.}} \href{http://dx.doi.org/10.1103/PhysRevLett.100.030602}{{\bf
  \bibinfo{volume}{100}}, \bibinfo{pages}{030602}}
  (\href{http://dx.doi.org/10.1103/PhysRevLett.100.030602}{\bibinfo{year}{2008}}).

\bibitem{flesch_cramer_08}
\bibinfo{author}{A.~Flesch}, \bibinfo{author}{M.~Cramer},
  \bibinfo{author}{I.~P. McCulloch}, \bibinfo{author}{U.~Schollw\"ock}, and
  \bibinfo{author}{J.~Eisert}, \bibinfo{title}{Probing local relaxation of cold
  atoms in optical superlattices},
  \bibinfo{journal}{\href{http://dx.doi.org/10.1103/PhysRevA.78.033608}{Phys.
  Rev. A}} \href{http://dx.doi.org/10.1103/PhysRevA.78.033608}{{\bf
  \bibinfo{volume}{78}}, \bibinfo{pages}{033608}}
  (\href{http://dx.doi.org/10.1103/PhysRevA.78.033608}{\bibinfo{year}{2008}}).

\bibitem{cazalilla_citro_review_11}
\bibinfo{author}{M.~A. Cazalilla}, \bibinfo{author}{R.~Citro},
  \bibinfo{author}{T.~Giamarchi}, \bibinfo{author}{E.~Orignac}, and
  \bibinfo{author}{M.~Rigol}, \bibinfo{title}{One dimensional bosons: From
  condensed matter systems to ultracold gases},
  \bibinfo{journal}{\href{http://dx.doi.org/10.1103/RevModPhys.83.1405}{Rev.
  Mod. Phys.}} \href{http://dx.doi.org/10.1103/RevModPhys.83.1405}{{\bf
  \bibinfo{volume}{83}}, \bibinfo{pages}{1405}}
  (\href{http://dx.doi.org/10.1103/RevModPhys.83.1405}{\bibinfo{year}{2011}}).

\bibitem{suppmat}
{\em \bibinfo{title}{\rm See Supplemental Material for details about the
  derivation of the emergent Hamiltonian, and for the explicit form of the
  boundary operators, for the quantum quench considered in this work}\/}.

\bibitem{peschel_eisler_09}
\bibinfo{author}{I.~Peschel} and \bibinfo{author}{V.~Eisler},
  \bibinfo{title}{Reduced density matrices and entanglement entropy in free
  lattice models},
  \bibinfo{journal}{\href{http://dx.doi.org/10.1088/1751-8113/42/50/504003}{J.
  Phys. A}} \href{http://dx.doi.org/10.1088/1751-8113/42/50/504003}{{\bf
  \bibinfo{volume}{42}}, \bibinfo{pages}{504003}}
  (\href{http://dx.doi.org/10.1088/1751-8113/42/50/504003}{\bibinfo{year}{2009}}).

\bibitem{calabrese_2005}
\bibinfo{author}{P.~Calabrese} and \bibinfo{author}{J.~Cardy},
  \bibinfo{title}{Evolution of entanglement entropy in one-dimensional
  systems},
  \bibinfo{journal}{\href{http://dx.doi.org/10.1088/1742-5468/2005/04/p04010}{J.
  Stat. Mech.}} \href{http://dx.doi.org/10.1088/1742-5468/2005/04/p04010}{{\bf
  \bibinfo{volume}{{\rm (2005)}}}, \bibinfo{pages}{P04010}}.

\bibitem{bloch08}
\bibinfo{author}{I.~Bloch}, \bibinfo{author}{J.~Dalibard}, and
  \bibinfo{author}{W.~Zwerger}, \bibinfo{title}{Many-body physics with
  ultracold gases},
  \bibinfo{journal}{\href{http://dx.doi.org/10.1103/RevModPhys.80.885}{Rev.
  Mod. Phys.}} \href{http://dx.doi.org/10.1103/RevModPhys.80.885}{{\bf
  \bibinfo{volume}{80}}, \bibinfo{pages}{885}}
  (\href{http://dx.doi.org/10.1103/RevModPhys.80.885}{\bibinfo{year}{2008}}).

\bibitem{bakr_gillen_09}
\bibinfo{author}{W.~S. Bakr}, \bibinfo{author}{J.~I. Gillen},
  \bibinfo{author}{A.~Peng}, \bibinfo{author}{S.~F\"{o}lling}, and
  \bibinfo{author}{M.~Greiner}, \bibinfo{title}{{A quantum gas microscope for
  detecting single atoms in a Hubbard-regime optical lattice}},
  \bibinfo{journal}{\href{http://dx.doi.org/10.1038/nature08482}{Nature}}
  \href{http://dx.doi.org/10.1038/nature08482}{{\bf \bibinfo{volume}{462}},
  \bibinfo{pages}{74}}
  (\href{http://dx.doi.org/10.1038/nature08482}{\bibinfo{year}{2009}}).

\bibitem{sherson_weitenberg_10}
\bibinfo{author}{J.~F. Sherson}, \bibinfo{author}{C.~Weitenberg},
  \bibinfo{author}{M.~Endres}, \bibinfo{author}{M.~Cheneau},
  \bibinfo{author}{I.~Bloch}, and \bibinfo{author}{S.~Kuhr},
  \bibinfo{title}{{Single-atom-resolved fluorescence imaging of an atomic Mott
  insulator}},
  \bibinfo{journal}{\href{http://dx.doi.org/10.1038/nature09378}{Nature}}
  \href{http://dx.doi.org/10.1038/nature09378}{{\bf \bibinfo{volume}{467}},
  \bibinfo{pages}{68}}
  (\href{http://dx.doi.org/10.1038/nature09378}{\bibinfo{year}{2010}}).

\bibitem{rigol_muramatsu_04sept}
\bibinfo{author}{M.~Rigol} and \bibinfo{author}{A.~Muramatsu},
  \bibinfo{title}{Universal properties of hard-core bosons confined on
  one-dimensional lattices},
  \bibinfo{journal}{\href{http://dx.doi.org/10.1103/PhysRevA.70.031603}{Phys.
  Rev. A}} \href{http://dx.doi.org/10.1103/PhysRevA.70.031603}{{\bf
  \bibinfo{volume}{70}}, \bibinfo{pages}{031603}}
  (\href{http://dx.doi.org/10.1103/PhysRevA.70.031603}{\bibinfo{year}{2004}}).

\bibitem{rigol_fitzpatrick_11}
\bibinfo{author}{M.~Rigol} and \bibinfo{author}{M.~Fitzpatrick},
  \bibinfo{title}{Initial-state dependence of the quench dynamics in integrable
  quantum systems},
  \bibinfo{journal}{\href{http://dx.doi.org/10.1103/PhysRevA.84.033640}{Phys.
  Rev. A}} \href{http://dx.doi.org/10.1103/PhysRevA.84.033640}{{\bf
  \bibinfo{volume}{84}}, \bibinfo{pages}{033640}}
  (\href{http://dx.doi.org/10.1103/PhysRevA.84.033640}{\bibinfo{year}{2011}}).

\bibitem{essler_fagotti_2016}
\bibinfo{author}{F.~H.~L. Essler} and \bibinfo{author}{M.~Fagotti},
  \bibinfo{title}{Quench dynamics and relaxation in isolated integrable quantum
  spin chains},
  \bibinfo{journal}{\href{http://dx.doi.org/10.1088/1742-5468/2016/06/064002}{J.
  Stat. Mech.}} \href{http://dx.doi.org/10.1088/1742-5468/2016/06/064002}{{\bf
  \bibinfo{volume}{{\rm (2016)}}}, \bibinfo{pages}{064002}}.

\bibitem{gramsch_rigol_12}
\bibinfo{author}{C.~Gramsch} and \bibinfo{author}{M.~Rigol},
  \bibinfo{title}{{Quenches in a quasidisordered integrable lattice system:
  Dynamics and statistical description of observables after relaxation}},
  \bibinfo{journal}{\href{http://dx.doi.org/10.1103/PhysRevA.86.053615}{Phys.
  Rev. A}} \href{http://dx.doi.org/10.1103/PhysRevA.86.053615}{{\bf
  \bibinfo{volume}{86}}, \bibinfo{pages}{053615}}
  (\href{http://dx.doi.org/10.1103/PhysRevA.86.053615}{\bibinfo{year}{2012}}).

\bibitem{murthy_srednicki_19}
\bibinfo{author}{C.~Murthy} and \bibinfo{author}{M.~Srednicki},
  \bibinfo{title}{{Relaxation to Gaussian and generalized Gibbs states in
  systems of particles with quadratic Hamiltonians}},
  \bibinfo{journal}{\href{http://dx.doi.org/10.1103/PhysRevE.100.012146}{Phys.
  Rev. E}} \href{http://dx.doi.org/10.1103/PhysRevE.100.012146}{{\bf
  \bibinfo{volume}{100}}, \bibinfo{pages}{012146}}
  (\href{http://dx.doi.org/10.1103/PhysRevE.100.012146}{\bibinfo{year}{2019}}).

\bibitem{gluza_19}
\bibinfo{author}{M.~Gluza}, \bibinfo{author}{J.~Eisert}, and
  \bibinfo{author}{T.~Farrelly}, \bibinfo{title}{{Equilibration towards
  generalized Gibbs ensembles in non-interacting theories}},
  \bibinfo{journal}{\href{http://dx.doi.org/10.21468/SciPostPhys.7.3.038}{SciPost
  Phys.}} \href{http://dx.doi.org/10.21468/SciPostPhys.7.3.038}{{\bf
  \bibinfo{volume}{7}}, \bibinfo{pages}{38}}
  (\href{http://dx.doi.org/10.21468/SciPostPhys.7.3.038}{\bibinfo{year}{2019}}).

\end{thebibliography}

\phantom{a}
\newpage
\setcounter{figure}{0}
\setcounter{equation}{0}
\setcounter{table}{0}

\renewcommand{\thetable}{S\arabic{table}}
\renewcommand{\thefigure}{S\arabic{figure}}
\renewcommand{\theequation}{S\arabic{equation}}
\renewcommand{\thepage}{S\arabic{page}}

\renewcommand{\thesection}{S\arabic{section}}

\onecolumngrid

\begin{center}

\setcounter{page}{1}

{\large \bf Supplemental Material:\\
Emergent eigenstate solution for generalized thermalization}\\

\vspace{0.3cm}

Yicheng Zhang$^{1}$, Lev Vidmar$^{2,3}$, and Marcos Rigol$^{1}$\\
$^1${\it Department of Physics, The Pennsylvania State University, University Park, Pennsylvania 16802, USA} \\
$^2${\it Department of Theoretical Physics, J. Stefan Institute, SI-1000 Ljubljana, Slovenia} \\ 
$^3${\it Department of Physics, Faculty of Mathematics and Physics, University of Ljubljana, SI-1000 Ljubljana, Slovenia}

\end{center}

\vspace{0.6cm}

\twocolumngrid

\label{pagesupp}

For the quantum quench considered in this work, one can write the operator $\hat{\cal M}(t)$ in Eq.~(1) in the main text as $\hat{\cal M}(t) = \hat{\cal H}(t) + \hat B(t)$, where $\hat{\cal H}(t)$ is the emergent Hamiltonian given by Eq.~(4) in the main text, and $\hat B(t)$ is the boundary operator. Using the notation in Eq.~(1), the emergent Hamiltonian has the form
\begin{eqnarray} \label{def_Hemet_long}
\hat{\cal H}(t) = \hat H_{\rm SF}^{(0)} - it\, \hat{\cal H}_1 \,,
\end{eqnarray}
where $\hat{\cal H}_1 = [\hat H_{\rm SF}^f, \hat H_{\rm SF}^{(0)}]$, and the boundary operator can be written as
\begin{eqnarray}
\hat B(t)=\sum_{n=2}^{\infty}\frac{(-it)^n}{n!}\hat {\cal B}_n \,,
\end{eqnarray}
where $\hat {\cal B}_2 = [\hat H_{\rm SF}^f, \hat{\cal H}_1]$ and
\begin{eqnarray}
\hat{\cal B}_{n+1} = [\hat H_{\rm SF}^f,\hat{\cal B}_n] \,,
\end{eqnarray}
for $n \geq 2$. The operator $\hat{\cal H}_1$ equals
\begin{eqnarray}
\hat{\cal H}_1=\sum_{l=-L/2+1}^{L/2-1}(\hat c^{\dagger}_{l+1} \hat c^{}_{l}-\hat c^{\dagger}_l \hat c^{}_{l+1})\,,
\end{eqnarray}
such that Eq.~(\ref{def_Hemet_long}) becomes Eq.~(4) in the main text.

Here we discuss the structure of the boundary terms $\hat {\cal B}_n$, which cause the deviation of the emergent eigenstate solution from the exact quantum dynamics~\cite{vidmar_iyer_17}. We first evaluate $\hat {\cal B}_2$:
\begin{equation}\label{eq:b2}
\hat {\cal B}_2=[\hat H_{\rm SF}^f,\hat{\cal H}_1]=2\left( \hat n_{-L/2+1}-\hat n_{L/2}\right),
\end{equation}
where $\hat n_{l}=\hat c^{\dagger}_l \hat c^{}_l$. Equation~\eqref{eq:b2} shows that $\hat {\cal B}_2$ is the difference of site-occupations at the boundary sites. For the next two ${\cal B}_n$ operators (for $n=3$ and 4), we have
\begin{equation}
\hat {\cal B}_3= [\hat H_{\rm SF}^f,\hat {\cal B}_2]= 2i\left(\hat j^{(1)}_{-L/2+1}+\hat j^{(1)}_{L/2-1}\right),
\end{equation}
and,
\begin{eqnarray}\label{B4}
\hat {\cal B}_4= [\hat H_{\rm SF}^f,\hat {\cal B}_3]&= &2\left[\hat h^{(2)}_{-L/2+1}+2\left(\hat n_{-L/2+1}-\hat n_{-L/2+2}\right)\right.\nonumber\\&&\quad\left.-\hat h^{(2)}_{L/2-2}+2(\hat n_{L/2-1}-\hat n_{L/2})\right],
\end{eqnarray}
where we define the generalized current operator $\hat j_l^{(m)}$, and the generalized kinetic energy operator $\hat h_l^{(m)}$, with support on $m+1$ sites, as:
\begin{eqnarray}
\hat j_l^{(m)}&=&(i\hat c^{\dagger}_{l+m} \hat c^{}_{l}+{\rm H.c.}), \\ \hat h_l^{(m)}&=&(\hat c^{\dagger}_{l+m} \hat c^{}_{l}+{\rm H.c.}),
\end{eqnarray}
We note that $\hat {\cal B}_3$ and $\hat {\cal B}_4$ only contain one-body operators that at most connect the boundary sites with their nearest and next-nearest neighbor sites, respectively.

Calculating $\hat {\cal B}_n$ for arbitrary values of $n$ involves computing the following commutators:
\begin{eqnarray} \label{commutator_h}
[\hat H_{\rm SF}^f,\hat j_l^{(m)}] &=& -i\left[\left(\hat h^{(m+1)}_{l}-\hat h^{(m+1)}_{l-1}\right) \right. \nonumber\\ &&\qquad\left.+ \left(\hat h^{(m-1)}_{l}-\hat h^{(m-1)}_{l+1}\right)\right], \\ \label{commutator_j}
[\hat H_{\rm SF}^f,\hat h_l^{(m)}] &=& i\left[\left(\hat j^{(m+1)}_{l}-\hat j^{(m+1)}_{l-1}\right) \right. \nonumber\\&& \quad\left. +\left(\hat j^{(m-1)}_{l}-\hat j^{(m-1)}_{l+1}\right)\right], 
\end{eqnarray}
and
\begin{equation}\label{commutator_n}
[\hat H_{\rm SF}^f,\hat n_l]= i\left(\hat j^{(1)}_{l}-\hat j^{(1)}_{l-1}\right)\,.
\end{equation}
One can see from Eqs.~(\ref{commutator_h})--(\ref{commutator_n}) that $\hat {\cal B}_n$ only contains one-body operators, and that their maximum support extends $n-1$ sites from the boundaries.

\end{document}